\documentstyle[12pt]{article}
\begin{document}
  \begin{titlepage}
  \begin{flushright}
SUSSEX-AST 93/2-1 \\
(February 1993)\\
  \end{flushright}
  \begin{center}
\Large
{\bf Perturbation Spectra from Intermediate Inflation} \\
\vspace{.3in}
\normalsize
\large{John D.~Barrow and Andrew R.~Liddle} \\
\normalsize
\vspace{.6 cm}
{\em Astronomy Centre,\\ University of Sussex, \\ Brighton BN1 9QH, U.K.}\\
\vspace{.6 cm}
\end{center}
\baselineskip=24pt
\begin{abstract}
\noindent
We investigate models of `intermediate' inflation, where the scale factor
$a(t)$ grows as $a(t) = \exp (A t^f)$, $0 < f < 1$, $A$ constant. These
solutions arise as exact analytic solutions for a given class of potentials
for the inflaton $\phi$. For a simpler class of potentials falling off as a
power of $\phi$ they arise as slow-roll solutions, and in particular they
include, for $f = 2/3$, the class of potentials which give the
Harrison--Zel'dovich spectrum. The perturbation spectral index $n$ can be
greater than unity on astrophysical scales. It is also possible to generate
substantial gravitational waves while keeping the scalar spectrum close to
scale-invariance; this latter possibility performs well when confronted with
most observational data.
\end{abstract}

\begin{center}
\vspace{1cm}
PACS numbers~~~98.80.Cq, 98.70.Vc\\
\end{center}
\end{titlepage}
\begin{center}
{\bf I.~~INTRODUCTION}
\end{center}

\vspace*{12pt}
Power-law and exponential inflationary universes are well studied. Exact
solutions exist in both cases and they are created by exponential and constant
scalar field potentials respectively \cite{INFL,PL}. Exact solutions can also
be found for `intermediate' inflationary universes in which the scale factor
expands as \cite{INTER}
\begin{equation}
\label{scal}
a(t) = \exp \left( A t^f \right) \quad ; \quad 0<f<1 \; \; , \; \; A>0 \quad
	{\rm constants}
\end{equation}
These models possess an array of interesting properties, particularly with
regard to the perturbation spectra they generate, and we shall here present
some of these properties.

The $k=0$ Friedmann universe containing a scalar field $\phi$ with potential
$V(\phi)$ obeys the equations ($8\pi G = c = \hbar = 1$)
\begin{eqnarray}
\label{eq1}
3 H^2 & = & \dot{\phi}^2/2 + V(\phi)\\
\label{eq2}
\ddot{\phi} + 3 H \dot{\phi} & = & - V'
\end{eqnarray}
where $H = \dot{a}/a$ is the Hubble parameter, and throughout overdots
indicate derivatives with respect to time and primes are derivatives with
respect to $\phi$.

An exact solution of Eqs.~(\ref{eq1}) and (\ref{eq2}) of the form of
Eq.~(\ref{scal}) exists \cite{INTER} with
\begin{equation}
V(\phi) = \frac{8 A^2}{(\beta+4)^2} \left[ \frac{\phi}{(2A\beta)^{1/2}}
	\right]^{-\beta} \, \left[ 6 - \frac{\beta^2}{\phi^2} \right]
\end{equation}
where $\beta = 4(f^{-1} - 1)$ and
\begin{equation}
\label{phit}
\phi = \left( 2A\beta t^f \right)^{1/2}
\end{equation}
For later use, we note that this allows one to write
\begin{equation}
\label{hphi}
H(\phi) = Af(2\beta A)^{\beta/4} \phi^{-\beta/2}
\end{equation}

The potential which gives rise to this solution is shown in figure 1. It
is negative for $0 < \phi^2 < \beta^2/6$, increases up to a maximum at
$\phi^2 = \beta(\beta+2)/6$ and then falls asymptotically to zero as
$1/\phi^{\beta}$ as $\phi \rightarrow \infty$. The solution exists anywhere
on this potential for $\phi>0$, by choice of the appropriate initial velocity
for $\phi$. In particular, to the left of the maximum of the potential the
field must be given a rapid velocity in order to cross the maximum and reach
the far side of the potential.

Note that although the solution is valid anywhere on the potential, it is not
always inflating. In fact, the condition for inflation ($\ddot{a} > 0$) is
only satisfied when $\phi^2 > \beta^2/2$, which guarantees that we must be in
the region of the potential where it is positive. For $\beta > 1$ ({\it ie} $f
< 4/5$), inflation can only occur beyond the maximum of the
potential\footnote{This raises an issue about the use of exact solutions ---
one must be wary to use them only when they serve as attractors for the
system. In this sense, the inflation that one can gain for $4/5 < f < 1$ to
the left of the maximum must be regarded as only a curiosity; the initial
condition which serves to fire the scalar field up the potential and over the
maximum is certainly not typical and the generic behaviour would be to roll
down the left hand side of the potential. Contrarywise, the inflation as the
potential rolls down to the right of the maximum is generic.}.

This form for $a(t)$ also arises when one solves the equations of motion in
the slow-roll approximation (see below) with a simple power-law potential
\begin{equation}
\label{SRPOT}
V(\phi) = \frac{48 A^2}{(\beta+4)^2} (2A\beta)^{\beta/2} \, \phi^{-\beta}
\end{equation}
Such a potential bears qualitative similarity to the exponential potentials of
power-law inflation. Here we note that the solutions for $\phi(t)$ and
$H(\phi)$ obtained for Eq.~(\ref{SRPOT}) in the slow-roll approximation are
identical to those obtained in the exact solution, Eqs.~(\ref{phit})
and (\ref{hphi}), and we shall exploit this later.

In some ways the slow-roll solution is more interesting than the exact
solution. In particular it arises from a much simpler form of the potential,
requiring only its asymptotic properties. It also possesses one rather curious
property, which is that no inflation occurs in the earliest stages of the
scalar rolling down the potential. While $\phi^2 < \beta^2/2$, the potential
is too steep. Only when the field reaches the asymptotic region of the
potential can inflation begin. If one assumes, following the usual `chaotic
inflation' philosophy, that the initial scalar energy is at the Planck
boundary, then inflation will always begin when the field reaches this value.

For intermediate inflation the slow-roll conditions become increasingly well
satisfied with time and so, like power-law inflation, there is no natural end
to inflation within the model. As with power-law inflation one expects this
state of affairs to be remedied by modifications to the potential which
create a minimum at a finite scalar field value. Intermediate inflation would
then arise only in a region of the potential. Another possibility
would be that, akin to Jordan--Brans--Dicke extended inflation \cite{EXINF}
for power-law inflation, intermediate inflation could arise in the conformal
frame of an extended inflation model and inflation could end via bubble
nucleation. Examples of this in scalar-tensor gravity theories have been given
by Barrow and Maeda \cite{BM} and Barrow \cite{B92}.

Before progressing, we formalise what we mean by the slow-roll approximation.
Noting that $2\dot{H} = - \dot{\phi}^2$, and assuming that during inflation
$\dot{\phi}$ never passes through zero so that we may divide by it,
substitution yields more useful forms of Eqs.~(\ref{eq1}) and (\ref{eq2}):
\begin{eqnarray}
\label{e1}
(H')^2 - \frac{3}{2} H^2 & = & - \frac{1}{2} \, V(\phi)\\
\label{e2}
\dot{\phi} & = & - 2 H'
\end{eqnarray}
In this formalism, it is possible to treat $H(\phi)$ as the fundamental
quantity to be chosen, rather than the more usual $V(\phi)$ \cite{LIDSEY}.

This formalism also allows a simple expression of the slow-roll conditions
to be made, one which is more fundamental than the version commonly seen
involving $V'/V$ and $V''/V$. Define slow-roll parameters $\epsilon$
and $\eta$ by\footnote{These definitions, as employed in \cite{REC}, differ
from those used in \cite{LL}.}
\begin{eqnarray}
\epsilon & \equiv & 3 \, \frac{\dot{\phi}^2/2}{V + \dot{\phi}^2/2} = 2 \left(
	\frac{H'}{H} \right)^2\\
\eta & \equiv & - 3 \, \frac{\ddot{\phi}}{3 H \dot{\phi}} = 2 \frac{H''}{H}
\end{eqnarray}
The slow-roll approximation is the assumption that both $\epsilon$ and
$|\eta|$ are small, and they correspond to the ability to neglect the first
term in Eq.~(\ref{e1}) and its derivative respectively. With these
definitions, the condition for inflation to occur, $\ddot{a} > 0$, is {\em
precisely} equivalent to $\epsilon < 1$.

\begin{center}
{\bf II.~~PERTURBATION SPECTRA FROM INTERMEDIATE INFLATION}
\end{center}

\vspace*{12pt}
It has long been recognised that inflation typically gives rise to a spectrum
of density perturbations close to the scale-invariant Harrison--Zel'dovich
form \cite{PERT}. Recent improved observations \cite{COBE} require that
deviations from this form be taken very seriously. Further, the possibility
that large-angle microwave background anisotropies may have contributions not
only from density perturbations, but also from gravitational wave modes
\cite{GRAV} which are excited during inflation, must be taken into
consideration \cite{LL,TENS}.

We shall simply quote standard results. The spectra of scalar and
transverse-traceless tensor perturbations are given \cite{DEF,LL} by the
expressions
\begin{eqnarray}
P_{{\cal R}}^{1/2} (k) & = & \left. \left( \frac{H^2}{2\pi |\dot{\phi}|}
	\right) \right|_{aH=k} = \left. \left( \frac{H^2}{4\pi |H'|} \right)
	\right|_{aH=k}\\
P_{g}^{1/2} (k) & = & \left. \left( \frac{H}{2\pi} \right) \right|_{aH=k}
\end{eqnarray}
where ${\cal R}$ is the perturbation in the spatial curvature. The expressions
on the right are to be evaluated when the comoving scale $k$ leaves the
horizon during inflation. These results hold to first order in the slow-roll
approximation, and we shall assume them throughout\footnote{Recently, the next
order corrections in slow-roll to these expressions have been calculated
\cite{LS}, but are rather cumbersome and will not be needed here.}. Useful
quantities are the spectral indices, which are scale-dependent in general.
They can be calculated from the above to first order in the slow-roll
parameters, as
\begin{eqnarray}
\label{NS}
n & \equiv & 1 + \frac{{\rm d} \ln P_{{\cal R}}}{{\rm d} \ln k} = 1 -
	4 \epsilon_* + 2 \eta_*\\
\label{NG}
n_g & \equiv & \frac{{\rm d} \ln P_g}{{\rm d} \ln k} = 2 \epsilon_*
\end{eqnarray}
where $n$ and $n_g$ are the scalar and gravitational wave spectral indices
respectively, and the star indicates that the slow-roll parameters should be
evaluated when the appropriate scale passes outside the horizon during
inflation (that is, at the value of the scalar field when the scale $k$
leaves the horizon during inflation). The flat Harrison--Zel'dovich
spectrum corresponds to $n=1$.

The most important quantity concerning the gravitational wave modes is the
extent to which they influence large-angle microwave anisotropies (on small
angular scales ($<2^0$), the gravitational waves would have been within the
horizon at the time of last scattering and their redshifting would have
reduced their significance). Thus, we decompose the temperature fluctuation
field into spherical harmonics
\begin{equation}
\frac{\Delta T}{T} (\theta, \phi) = \sum_{l,m} a_{lm} Y^l_m (\theta, \phi)
\end{equation}
In a given inflationary model, it is simple to calculate the contribution to
the variances of the $a_{lm}$, which are $m$-independent due to rotational
invariance and can be denoted $\Sigma_l^2$. The general expressions are given
as integrals over the power spectra, but in the slow-roll approximation,
the relative contribution $R_l$ of scalar and tensor modes is well
approximated \cite{LL} by the simple expression
\begin{equation}
R_l \equiv \frac{\Sigma_l^2 ({\rm tensor})}{\Sigma_l^2 ({\rm scalar})} = 12.4
	\epsilon_l
\end{equation}
where $\epsilon_l$ indicates that the slow-roll parameter is to be evaluated
at the scale corresponding to the $l$-th multipole, $k = l H_0/2$.

Details of all the above expressions can be found in \cite{LL}. If the
slow-roll conditions are well satisfied, then the spectrum must be close to
flat and the contribution from gravitational waves to the COBE signal must be
small.

The benefit of the $H(\phi)$ behaviour being the same for both the exact
solution and slow-roll solution is immediately apparent --- they give rise to
the same density perturbation spectra. Indeed, one need not ask what the
origin of the particular form of $H(\phi)$ was. One can compute the slow-roll
parameters during inflation; these are given by
\begin{eqnarray}
\epsilon & = & \frac{\beta^2}{2\phi^2}\\
\eta & = & \beta ( 1+ \beta/2) \frac{1}{\phi^2}
\end{eqnarray}
They therefore possess a mild scale-dependence, whereas in power-law inflation
from an exponential potential they would be constant in time.

The amplitude of the scalar spectrum will depend on the amplitude of the
potential, and is to be fixed by observations such as COBE. More interesting
is the scale-dependence of the scalar spectrum. This is given from
Eq.~(\ref{NS}) by
\begin{equation}
\label{SPEC}
n = 1 - \frac{\beta(\beta-2)}{\phi^2}
\end{equation}
Recall that $\beta = 2$ corresponds to $f = 2/3$.

This spectrum offers properties which are unusual in inflation, where the
typical behaviour (exhibited by polynomial chaotic inflation, power-law
inflation, natural inflation, extended inflation, etc) is for a spectrum with
$n < 1$, so that a COBE normalised spectrum has reduced small-scale power
compared to a similarly normalised scale-invariant spectrum. However, provided
$0 < \beta < 2$ ($1 > f > 2/3$), the spectrum Eq.~(\ref{SPEC}) offers a value
of $n$ greater than 1, albeit with a scale-dependence we address below.

As one expects, the limit of exponential expansion ($f \rightarrow 1$)
gives rise to a flat Harrison--Zel'dovich spectrum, though were one to compute
the amplitude one would find it diverging in this limit. Much more interesting
is the case $f = 2/3$. This too offers a flat spectrum, but now with finite
(in fact freely selectable) amplitude. The potential therefore corresponds to
that giving (in the slow-roll approximation) an exactly flat spectrum.
Although this potential has arisen before on general grounds \cite{HB,REC}, we
are not aware of it having been identified as that giving intermediate
inflation before.

The gravitational spectral index is only of modest interest; more significant
is the relative contribution $R_l$ of tensors and scalars to the COBE signal.
This is just
\begin{equation}
R_l \simeq 12.4 \epsilon_l = 6.2 \beta^2/\phi^2
\end{equation}
where $\phi$ is evaluated at the time the scale corresponding to the $l$-th
multipole leaves the horizon. In general, the relative amplitude is
related to the scalar spectral index, by
\begin{equation}
n = 1 - \frac{\beta -2}{6.2 \beta} R_l
\end{equation}
where $n$ is given on the scale corresponding to the $l$-th
multipole\footnote{Compare with the power-law inflation result $n = 1 -
R_l/6.2$. In that case $n$ and $R_l$ are both scale-independent.}.

For typical choices of $\beta$, if we are on the part of the potential
where the spectrum is close to flat, there are few gravitational waves, in
keeping with the slow-roll conditions. However, for the exact
Harrison--Zel'dovich case $\beta=2$, the relative contribution depends simply
on the value of $\phi$ when the relevant scales leave the horizon. In terms of
the slow-roll parameters, one is arranging $\eta_* = 2\epsilon_*$, without
requiring that either separately be small. As $\phi$ is only constrained by
the inflation condition $\phi^2 > \beta^2/2$, the gravitational waves can in
fact be the dominant contributor to COBE. This is also possible for other
choices of $\beta$ close to 2.

We get departures from `almost' scale invariance with negligible gravitational
waves only for small values of $\phi$. Thus, for cosmological interest, one
needs to be on this part of the potential as large-scale structure scales
leave the horizon during inflation. With standard reheating, this will
correspond to around 60 $e$-foldings from the end of inflation. The number of
$e$-foldings of intermediate inflation between scalar field values $\phi_1$
and $\phi_2$ is given by
\begin{equation}
N(\phi_1,\phi_2) = - \frac{1}{2} \int_{\phi_1}^{\phi_2} \frac{H}{H'} {\rm d}
	\phi = \frac{1}{2\beta} \left( \phi_2^2 - \phi_1^2 \right)
\end{equation}
Without knowing how inflation ends, one cannot draw any further conclusions
from this, because the number of $e$-foldings to complete inflation is
unknown.

Suppose that we assume inflation begins at the smallest $\phi$ value which
permits it, so $\phi_1^2 = \beta^2/2$. As discussed after Eq.~(\ref{SRPOT}),
this seems very reasonable in the slow-roll case. The spectral index and
gravitational wave contributions can now be expressed in terms of the number
of $e$-foldings $N_b$ which have passed since the {\em beginning} of inflation
\begin{eqnarray}
n & = & 1 - \frac{2\beta - 4}{4 N_b + \beta}\\
R_l & = & \frac{12.4 \beta}{4 N_b + \beta}
\end{eqnarray}
One expects that the total amount of inflation must exceed the 60 $e$-foldings
mentioned above by a factor of at least a few\footnote{Of course the total
amount of inflation could be enormous, in which case we are guaranteed the
slow-roll limit.} (one often sees 70 as the minimum number required to solve
the flatness problem). This sets upper limits on the $|n-1|$ and $R_l$ as a
function of $\beta$. For example, for the Harrison--Zel'dovich spectrum case
$\beta = 2$ this would imply that the gravitational wave contribution to COBE
is subdominant ($R_l < 0.6$) though certainly not insignificant.

This also allows us to address the scale-dependence of the spectral index ---
would one be able to see deviations from power-law behaviour across observable
scales? Potentially observable scales, those in the linear regime at the
present, stretch from about $3000 h^{-1}$ Mpc down to $8h^{-1}$ Mpc, which
corresponds to about 6 $e$-foldings of scale. Putting this into the
expressions above, we see that the typical scale-dependence of $n$ is rather
limited, even in cases where the deviation from $n=1$ is dramatic (notably
large $\beta$).

\newpage
\begin{center}
{\bf III.~~DISCUSSION}
\end{center}

\vspace*{12pt}
We have investigated models of intermediate inflation. Intermediate inflation
arises as the slow-roll solution to potentials which fall off asymptotically
as a power law in $\phi$, and can be modelled by an exact cosmological
solution. This model bears many qualitative similarities to power-law
inflation: like power-law inflation, there is no natural end to inflation and
a mechanism must be introduced in order to bring inflation to an end. Also, as
with power-law inflation, intermediate inflation offers the possibility of
density perturbation and gravitational wave spectra which differ significantly
from the usual inflationary prediction of a nearly flat spectrum with
negligible gravitational waves.

In particular, two interesting types of behaviour can be obtained which do not
arise in traditional inflationary models:
(1) It is possible for the spectral index to exceed unity over large scale
structure scales;
(2) Intermediate inflation contains the class of models which generate the
exact Harrison--Zel'dovich spectrum (in the slow-roll approximation). Within
this class, there exist models which produce substantial gravitational waves
despite the flatness of the spectrum.

This latter case is of particular interest, because it completes a square of
possible inflationary predictions for the tilt of the density spectrum and the
influence of gravitational waves. This is illustrated in Table 1. This
model would perform well on confrontation with large-scale structure data in a
cold dark matter model, as the gravitational wave contribution could explain
the unexpectedly large amplitude of the COBE result, especially should the
true result prove to lie towards the lower end of the COBE range. The most
troublesome data would be the clustering data such as the APM survey
\cite{APM}; one would have to resort to astrophysical effects ({\it eg}
cooperative galaxy formation \cite{COOP}) in order to explain this.

%
%

However, we must remark that the models which give these deviations from the
usual predictions are rather special, in that we are relying on the minimum
amount of inflation occurring so that large-scale structure scales cross the
horizon at a time when the slow-roll conditions were still not well obeyed.
This corresponds to a non-trivial constraint on when modifications to the
potential entered to end inflation. However, this is still quite a strong
result because it implies that obtaining a nearly scale-invariant spectrum
with significant gravitational waves, while possible in principle, is very
difficult in practice because the scale-invariance in the density perturbation
spectrum implies that the relative contribution from gravitational waves falls
as inflation proceeds.

\begin{center}
{\bf ACKNOWLEDGEMENTS}
\end{center}

\vspace*{12pt}
JDB is supported by the Leverhulme Trust and the Royal Society. ARL is
supported by the SERC, and acknowledges the use of the Starlink computer
system at the University of Sussex.
\frenchspacing

\newpage
\vspace*{24pt}
\nonfrenchspacing
\small
\begin{tabular}{|c||c|c|}
\hline
& Small gravitational-wave & Large gravitational-wave\\
& contribution to COBE & contribution to COBE\\
\hline \hline
Nearly flat spectra & Polynomial chaotic inflation \cite{LINDE} & Intermediate
inflation ($f \simeq 2/3$) \cite{INTER}\\
\hline
Tilted spectra & Natural inflation \cite{NAT} & Power-law inflation (small
power) \cite{PL}\\
\hline
\end{tabular}

\vspace*{24pt}
\noindent
{\em Table 1}\\
Examples of inflationary models giving rise to different types of prediction
for the perturbation spectra.

\vspace{1cm}

\noindent
{\em Figure 1}\\
The potential which gives exact intermediate inflation, for sample parameter
choices $\beta = 2$ ($f=2/3$) and $A=1$. Generically, the maximum is located
at $\phi^2 = \beta(\beta+2)/6$.

\end{document}